\documentclass[11pt,a4paper]{article}

\usepackage[utf8]{inputenc}
\usepackage[T1]{fontenc}
\usepackage{lmodern}
\usepackage{microtype}
\usepackage{geometry}
\usepackage{setspace}
\usepackage{hyperref}
\hypersetup{
  colorlinks=true,
  linkcolor=black,
  citecolor=blue,
  urlcolor=blue,
  pdfauthor={},
  pdftitle={JASDA: Introducing Job-Aware Scheduling in Scheduler-Driven Job Atomization}
}
\usepackage{graphicx}
\usepackage{caption}
\usepackage{subcaption}
\usepackage{xcolor}
\usepackage{amsmath,amssymb}
\usepackage{booktabs}
\usepackage{enumitem}
\usepackage{algorithm}
\usepackage{algpseudocode}
\usepackage{orcidlink}
\usepackage{authblk}
\usepackage{array,tabularx}

\usepackage{makecell}
\usepackage{ragged2e}

\newcommand{\SelectBestCompatibleVariants}{\textsc{Select\-Best\-Compatible\-Variants}}

\usepackage[acronym]{glossaries}
\makenoidxglossaries

\newacronym{SJA}{SJA}{Scheduler-Driven Job Atomization}
\newacronym{JASDA}{JASDA}{Job-Aware Scheduling Algorithm}
\newacronym{MIG}{MIG}{Multi-Instance GPU}
\newacronym{QoS}{QoS}{Quality of Service}
\newacronym{GPU}{GPU}{Graphical Processing Unit}
\newacronym{JCT}{JCT}{Job Completion Time}
\newacronym{MPS}{MPS}{Multi-Process Service}
\newacronym{ILP}{ILP}{Integer Linear Programming}
\newacronym{TRP}{TRP}{Temporal Resource Profile}
\newacronym{FMP}{FMP}{Functional Memory Profile}
\newacronym{DP}{DP}{Dynamic Programming}
\newacronym{WIS}{WIS}{Weighted Interval Scheduling}

\title{\textbf{JASDA: Introducing Job-Aware Scheduling in Scheduler-Driven Job Atomization}}
\author[1]{Michal Konopa\,\orcidlink{0000-0003-1694-1529}}
\author[1]{Jan Fesl\,\orcidlink{ 0000-0001-7192-4460}}
\author[1]{Ladislav Ber{\'a}nek\,\orcidlink{0000-0001-5004-0164}}

\affil[1]{Department of Data Science and Computer Systems, Faculty of Agriculture and Technology, University of South Bohemia, Czech Republic\\
\texttt{konopm05@fzt.jcu.cz}, \texttt{jfesl@fzt.jcu.cz}, \texttt{beranek@fzt.jcu.cz}}
\date{October 2025}

\begin{document}
\maketitle

\begin{abstract}
\noindent
The increasing complexity and temporal variability of workloads on \gls{MIG}-enabled \gls{GPU}s challenge the scalability of traditional centralized scheduling. 
Building upon the \gls{SJA} concept, this paper introduces \gls{JASDA}---a novel paradigm that extends \gls{SJA} from a largely centralized scheduling model toward a fully decentralized negotiation process. 

In \gls{JASDA}, jobs actively generate and score feasible subjobs in response to scheduler-announced execution windows, while the scheduler performs policy-driven clearing that balances utilization, fairness, and temporal responsiveness. This bidirectional, iterative interaction embeds feedback, calibration, and probabilistic safety directly into the scheduling loop, enabling adaptive and transparent decision-making.

By coupling principles from auction theory and online optimization with the temporal granularity of GPU workloads, \gls{JASDA} provides a scalable foundation for market-aware and fairness-driven resource management---bridging theoretical scheduling models with practical deployment in modern \gls{MIG}-enabled environments relevant to Artificial Intelligence and Agriculture~4.0.
\end{abstract}

\vspace{0.5em}

\section{Introduction}
\gls{MIG} architectures~\cite{NVIDIA2025MIG} enable the physical partitioning of a single GPU into multiple isolated instances (“slices”), thereby supporting multi-tenancy---simultaneous sharing of a GPU by multiple jobs. Nevertheless, as workloads become increasingly heterogeneous and dynamic (e.g., AI model training and inference, data analytics, or domain-specific computational pipelines for Agriculture~4.0), classical centralized schedulers struggle to reduce idle time and resource underutilization while maintaining \gls{QoS} guarantees without incurring excessive job preemption, reconfiguration, or migration overhead.

The \gls{SJA} concept redefines job execution as a stream of subjobs (atoms) that can be scheduled opportunistically. In this scheduling model, jobs already respond to scheduler-announced execution windows by generating feasible subjobs; however, the overall scheduling logic remains largely centralized---the scheduler alone performs global evaluation and allocation. Elevating jobs from passive recipients to \emph{decision-capable agents}, capable of expressing multiple feasible variants and their local preferences, represents the natural next step.

This paper introduces \gls{JASDA}, a job-aware scheduling algorithm in which: \\
(i) in each scheduling iteration, the scheduler announces a single available execution window---a free time interval---on a specific slice, \\
(ii) jobs respond with a list of variant proposals (i.e., subjobs annotated with self-evaluated scores), and \\
(iii) the scheduler then evaluates and clears that window by selecting a non-conflicting subset of jobs (ensuring that their execution intervals do not overlap) according to a combined job–system criterion. \\ 
The contribution of this concept paper is threefold:
\begin{enumerate}[leftmargin=1.5em]
  \item \textbf{Interaction cycle.} We formalize the \emph{\gls{JASDA} interaction cycle}---a bidirectional, preference-driven loop atop \gls{SJA}---and its core principles (Section~\ref{sec:conceptual_framework}).
  \item \textbf{Per-window clearing algorithm.} We specify a lightweight, \gls{WIS}-based selection routine with optimal per-window clearing, its scoring model, and complexity (Section~\ref{sec:algorithm}).
  \item \textbf{Illustration and discussion.} We provide an illustrative example and discuss expected benefits, limitations, and design implications, laying the groundwork for a follow-up experimental study (Section~\ref{sec:illustrative_example}).
\end{enumerate}

\paragraph{Relation to SJA.}
\gls{JASDA} builds directly on the \textit{\gls{SJA}} concept~\cite{konopa2025sja}, which established a bidirectional offer--reply interaction between the scheduler and jobs. In \gls{SJA}, jobs are decomposed into eligible subjobs using \gls{FMP} and accepted only if they satisfy a predefined risk bound 
$\Pr(\max_t \text{RAM}(t) \le C) \ge 1-\epsilon$. 
\gls{JASDA} \textbf{extends} this concept with a concrete \emph{variant-bidding layer} and an \emph{optimal per-window clearing mechanism}. Jobs propose lists of eligible variants along with self-assessed utility values for each variant, while the scheduler aggregates these lists, evaluates combined system–job scores for all variants, and performs an optimal selection of subjobs within each active scheduling window.

\paragraph{Paper scope.} This paper is a concept paper. It primarily focuses on the conceptual and design rationale, followed by an algorithmic formulation of the scheduling approach. Implementation details and quantitative analysis are deferred to a subsequent journal paper.

\section{Motivation and Background}
Recent works such as \textit{MISO}~\cite{Li2022}, \textit{MIGRator}~\cite{Wang2024MIGRator}, and \textit{Pollux}~\cite{Qiao2021Pollux} have explored GPU slicing, dynamic reconfiguration, and adaptive cluster-level scheduling for \gls{MIG}-enabled systems. A more detailed analysis of their limitations has already been provided in \textit{\gls{SJA}}~\cite{konopa2025sja}, which introduced the first interactive scheduling concept enabling jobs to act as autonomous agents that independently respond to scheduler-announced execution windows. Building on this foundation, the present work extends the concept toward an explicit, bidirectional decision layer--\textit{\gls{JASDA}}--in which jobs autonomously generate and evaluate multiple execution variants in response to scheduler offers. This formulation naturally aligns with the principles of auction- and market-based decision-making, providing a bridge between interactive job autonomy and economically inspired resource allocation models.

Auction- and market-based scheduling approaches have proven highly effective in distributed decision-making and resource allocation. They rely on the key principle of iterative interaction between the \textit{bidding} and \textit{clearing} phases: agents express their local preferences and utilities through bids, while a centralized or distributed auctioneer evaluates these bids and selects those that maximize the overall system utility. The classical auction algorithm by Bertsekas~\cite{BERTSEKAS1991707}, which laid the theoretical foundation for this approach, models resource allocation as an iterative bidding process that converges toward the optimal assignment. Wellman et~al.~\cite{WELLMAN2001271} further extended this idea to a decentralized scheduling paradigm, where autonomous agents negotiate resource allocation through local auction protocols without global synchronization. Such market-based methods have since been widely applied in cloud and edge computing~\cite{Kumar2021SurveyEdge} to achieve economically efficient and fair allocation of virtualized resources. A notable example is \textit{Themis}~\cite{Mahajan2020}, which introduced a bidding-based scheduling framework for GPU clusters that balances finish-time fairness and overall efficiency through auction-style negotiation between jobs and the scheduler. However, these approaches still treat individual jobs as indivisible, monolithic entities and overlook the finer-grained resource elasticity offered by MIG architectures. The \textit{\gls{SJA}} concept~\cite{konopa2025sja} represents a new paradigm bridging these two worlds through bidirectional negotiation between the scheduler and jobs, while the proposed \textit{\gls{JASDA}} concept further extends this paradigm by allowing jobs to generate and evaluate multiple feasible execution variants---thereby embedding auction-based decision logic directly into the runtime operation of a \gls{MIG}-aware scheduler.

\subsection{From Auction Mechanisms to Per-Slice Bidding in MIG Scheduling}
Although auction-based scheduling approaches have been extensively studied in distributed systems and cloud environments, their adaptation to GPU-enabled systems requires a conceptual shift in how bidding and clearing mechanisms can be effectively mapped onto the granularity of individual slices and time windows. In conventional cloud  or multi-robot auctions, agents typically compete for indivisible computational resources such as CPU cores, with each bid representing a scalar measure of utility. In contrast, \gls{MIG}-enabled systems expose a multidimensional allocation space encompassing both spatial (slice capacity) and temporal (available window length) dimensions, which inherently leads to a richer bidding landscape.

Within this model---an auction-inspired scheduling framework for \gls{MIG} environments---each job can be viewed as an agent capable of evaluating the utility of different variants of its execution. Each variant corresponds to a specific subjob, defined by its resource requirements (e.g., slice capacity, runtime duration, etc.). The job assigns a utility score to each variant, representing how well the variant aligns with its internal objectives and constraints, such as expected completion time or adherence to \gls{QoS} requirements. The evaluated variants are then submitted as bids to the scheduler, which acts as a market-clearing entity and selects a combination of variants that maximizes a global objective function---typically reflecting overall resource utilization, fairness, or makespan minimization.

In summary, the \textit{\gls{JASDA}} concept views job scheduling in \gls{MIG}-enabled systems implementing \textit{\gls{SJA}} as an auction in which individual jobs bid with lists of their feasible execution variants for specific open windows in the schedule. This perspective allows scheduling to be interpreted as a market-clearing process, where local job preferences translate into global system efficiency. It provides a natural theoretical framework that links heuristic scheduling strategies with auction-theoretic principles and enables formal reasoning about their efficiency, stability, and fairness.

\section{Conceptual Framework of JASDA}
\label{sec:conceptual_framework}

\textbf{Core principles.}
\gls{JASDA} is based on three complementary principles that together define its adaptive and self-organizing nature.

\begin{description}[leftmargin=2cm, labelindent=0cm, style=nextline]
    \item[\textbf{(1) Bidirectional interaction.}] 
    The scheduler advertises available time--capacity windows, and jobs respond with eligible subjob proposals that can safely fill these gaps. 
    This feedback loop enables adaptive, context-aware scheduling decisions instead of static, one-sided allocation.

    \item[\textbf{(2) Variant-based job modeling.}] 
    Each job can propose several execution variants, representing alternative subjob configurations defined by duration, capacity, and start time. 
    This exposes flexibility on the job side and allows the scheduler to better utilize fragmented \gls{MIG} resources.

    \item[\textbf{(3) Policy-driven decision integration.}] 
    The scheduler selects subjobs according to a policy that balances job preferences with system-wide objectives such as utilization, fairness \emph{(including temporal fairness; see Section~\ref{subsec:age_fairness})}, or slack minimization. This ensures system efficiency while respecting job-level priorities.
\end{description}

Together, these principles enable a self-organizing and adaptive scheduling process that can efficiently fill fine-grained time--capacity gaps across heterogeneous \gls{MIG} slices while maintaining alignment between local (job-side) and global (system-wide) objectives.

\vspace{0.5em}
\noindent \textbf{High-level operation.}
The \gls{JASDA} scheduling loop proceeds in discrete iterations, each consisting of five main steps summarized below. This cycle repeats continuously as jobs arrive and resources become available.

\begin{enumerate}[leftmargin=*, label=\textbf{Step~\arabic*:}, itemsep=0.4em]
  \item \textbf{Window announcement.}  
  The scheduler selects an available time–capacity window on a specific slice and broadcasts its parameters to all jobs in the waiting queue.

  \item \textbf{Job-side variant generation.}  
   Each job evaluates the announced window and autonomously decides whether to respond by generating one or more \emph{eligible} subjob variants. A subjob is considered eligible if it satisfies all temporal and capacity constraints of the announced window, as well as any additional technical or policy-related eligibility conditions defined by the system. Jobs that cannot produce an eligible subjob simply remain silent, preserving the agent-driven nature of the interaction and avoiding any scheduler-side prefiltering.
   
  \item \textbf{Bid submission.}  
  Jobs submit their variants as bids, each associated with a \emph{local utility score} representing expected benefit or performance gain.

  \item \textbf{Scheduler clearing.}  
  The scheduler evaluates all bids using a combined scoring function that balances job utilities and system-level metrics, and selects the best-fitting subjob or sequence of temporally non-overlapping subjobs.

  \item \textbf{Commit and advance.}  
  The selected subjob(s) are committed to the schedule, resources are updated, and the process advances to the next iteration.
\end{enumerate}

This high-level cycle transforms scheduling into an ongoing negotiation between the scheduler and jobs, where both sides dynamically adapt their behavior to evolving resource conditions.

\subsection{Step 1: Window Announcement}
\label{subsec:announcement_phase}
In each iteration, the scheduler identifies one candidate time–capacity windows across the available \gls{MIG} slices and announces it to all jobs. Each announced window $w^\ast=(s_k, c_k, t_{\min}, \Delta t)$ corresponds to a contiguous idle region on slice $s_k$, where:
\begin{itemize}[leftmargin=1.5em]
    \item $s_k$ -- identifier of the \gls{MIG} slice;
    \item $c_k$ -- memory capacity of slice $s_k$;
    \item $t_{\min}$ -- start time of the earliest available idle segment;
    \item $\Delta t$ -- duration of the time window during which the slice remains available.
\end{itemize}
Intuitively, $w^\ast$ represents a potential scheduling opportunity --- a temporary \emph{gap} in which one or more subjobs can be safely executed without preemption or capacity conflict.

\paragraph{Window Selection Policy.}
The efficiency of the scheduling depends critically on which free scheduling window is selected for job bidding. From a practical perspective, the time required for subjob generation and scoring must also be taken into account. Therefore, it is generally preferable to announce window whose start time is near enough to ensure responsiveness, yet not so imminent that jobs lack sufficient time for bid preparation. While more elaborate selection criteria can be envisioned---such as earliest-finish or slack-aware heuristics, or even adaptive announcement strategies---the present concept paper focuses on the fundamental job–scheduler interaction and does not attempt a comparative evaluation of alternative policies. A systematic study of window selection strategies is left for future work (see Section~\ref{subsec:challenges}).

\subsection{Step 2: Job-Side Variant Generation}
\label{subsec:variant_generation}
Upon receiving the announcement, each job \(J_i\) evaluates the announced window \(w^\ast\) and determines whether it can propose one or more \emph{eligible subjobs}. 
A subjob is considered \emph{eligible} if it satisfies all temporal and capacity constraints of the announced window and complies with any additional technical or policy-related requirements defined by the system (e.g., job–slice compatibility or administrative priorities).

\paragraph{\gls{TRP} and \gls{FMP}.}
A \emph{\gls{TRP}} for job~$J_i$ is a probabilistic model of its time–varying resource demand over its execution; it captures the distribution of temporal demand trajectories, including warm-up phases, steady-state intervals, and transient bursts. A \emph{\gls{FMP}} is a TRP specialized to device memory usage. 
In \gls{JASDA}, \gls{TRP}s and \gls{FMP}s serve two fundamental roles:
\begin{itemize}[leftmargin=1.5em,topsep=2pt,itemsep=2pt]
    \item deriving the \emph{predicted duration} $\tilde{\Delta t}_i$ of each subjob variant proposed into an announced window, and 
    \item verifying \emph{probabilistic safety} (“safe-by-construction”) with respect to the slice capacity~$c_k$ over the predicted execution interval.
\end{itemize}
The concepts of \gls{TRP} and \gls{FMP} originate from the \textit{\gls{SJA}} concept~\cite{konopa2025sja}, where they were introduced as compact probabilistic descriptors enabling safe and efficient temporal placement of subjobs.

\paragraph{Formal definition of a variant.}
For an announced window $w^\ast$, a variant proposed by job~$J_i$ is a tuple
\[
v_{i,k,w^\ast} \;=\; \big(s_k,\; t_{\mathrm{start}},\; \tilde{\Delta t}_i,\; \mathrm{TRP}_i\big).
\] where:
\begin{itemize}[leftmargin=1.5em]
    \item $s_k$ -- identifier of the \gls{MIG} slice associated with the announced window $w^\ast$ (capacity $c_k$ is given by $w^\ast$)
    \item $t_{\mathrm{start}}$ -- planned start time chosen within the announced window, i.e., $t_{\mathrm{start}}\in [t_{\min},\, t_{\min}{+}\Delta t]$,
    \item $\tilde{\Delta t}_i$ -- \emph{predicted duration} of the proposed subjob, derived from the job’s \gls{TRP},
    \item $\mathrm{TRP}_i$ -- a compact \emph{\gls{TRP}} descriptor for the variant (typically an \emph{FMP} for memory), capturing the expected time–varying resource usage needed for variant score evaluation and safety checks.
\end{itemize}
The predicted execution interval is
\[
I(v_{i,k,w^\ast})\;=\;[\,t_{\mathrm{start}},\, t_{\mathrm{start}}{+}\tilde{\Delta t}_i\,].
\]

Each variant therefore specifies a feasible execution interval within the announced window and can be uniquely identified by its tuple parameters. 

A job may generate multiple alternative eligible variants for a given announced window, each associated with a local utility score $h(v_{i,k,w^\ast})$. The overall score for each variant is then determined by the scheduler based on its local utility $h(v_{i,k,w^\ast})$ and system-wide optimization criteria.

\subsection{Step 3: Bid Submission}
Each job forms a portfolio of eligible variants \(\mathcal{V}_i = \{v_{i,k,w^\ast}\}\), which constitutes its bid set for the current iteration. The bids and their local utility scores are sent to the scheduler, representing how attractive the announced window is from the job’s perspective. This design mirrors the logic of market-based mechanisms, where self-interested agents reveal their local preferences through scored bids.

\subsection{Step 4: Scheduler Clearing and Selection}
After collecting all bids from jobs, the scheduler evaluates each proposed variant using a combined \emph{scoring function} that balances job-side and system-side objectives. Each variant is assigned a scalar score reflecting its overall utility, taking into account both the job’s internal preference (e.g., completion time, QoS targets, or energy cost) and the system’s global optimization goals (e.g., utilization, fairness, fragmentation minimization). The detailed formulation of the scoring model and its normalization scheme is presented in Section~\ref{subsec:scoring_model}.

Based on these scores, the scheduler selects the variant---or a compatible subset of variants---that maximizes the aggregate value while ensuring temporally non-overlapping allocations on each slice. This clearing step yields a locally optimal assignment for the current scheduling iteration and updates the system state for the next announcement cycle.

\subsection{Step 5: Commit and Advance}
The selected variants are committed to the execution plan. The scheduler updates its time–capacity map to reflect new allocations, while jobs record the outcome of their bids for monitoring and evaluation purposes. The cycle then advances, potentially triggering a rolling “repack” or defragmentation step when residual gaps become too small for further allocation.

\subsection{Comparison with Existing MIG Scheduling Approaches}
Table~\ref{tab:comparison} summarizes conceptual differences between \gls{JASDA} and representative \gls{MIG}-aware schedulers. While earlier systems such as MISO~\cite{Li2022}, MIGRator~\cite{Wang2024MIGRator}, and Pollux~\cite{Qiao2021Pollux} focus on static or reactive optimization, \gls{JASDA} introduces a continuous, market-inspired process in which jobs actively generate and score their execution variants.

\begin{table}[t]
\centering
\caption{Conceptual comparison between JASDA and representative MIG-aware schedulers.}
\label{tab:comparison}
\begin{tabular}{@{}p{3.1cm}p{3.6cm}p{3.6cm}p{3.8cm}@{}}
\toprule
\textbf{Aspect} &
\textbf{MISO / MIGRator} &
\textbf{Pollux / HiveD} &
\textbf{JASDA (proposed)} \\
\midrule

\makecell[{{p{3.1cm}}}]{Scheduling model} &
\makecell[{{p{3.6cm}}}]{\RaggedRight Static or reactive; performance- / ILP-driven reconfiguration.} &
\makecell[{{p{3.6cm}}}]{\RaggedRight Cluster-level optimization; fairness / co-adaptive training.} &
\makecell[{{p{3.8cm}}}]{\RaggedRight Cyclic, bidirectional scheduling with per-iteration negotiation.} \\

\makecell[{{p{3.1cm}}}]{Granularity} &
\makecell[{{p{3.6cm}}}]{\RaggedRight Job or slice-level layouts.} &
\makecell[{{p{3.6cm}}}]{\RaggedRight Multi-device / VC layouts.} &
\makecell[{{p{3.8cm}}}]{\RaggedRight Per-slice, per-window (gap-focused) decisions.} \\

\makecell[{{p{3.1cm}}}]{Job participation} &
\makecell[{{p{3.6cm}}}]{\RaggedRight Passive (scheduled by the system).} &
\makecell[{{p{3.6cm}}}]{\RaggedRight Passive.} &
\makecell[{{p{3.8cm}}}]{\RaggedRight Active: jobs propose multiple scored variants (bids).} \\

\makecell[{{p{3.1cm}}}]{Adaptivity} &
\makecell[{{p{3.6cm}}}]{\RaggedRight Reconfiguration / migration.} &
\makecell[{{p{3.6cm}}}]{\RaggedRight Policy-driven at cluster scale.} &
\makecell[{{p{3.8cm}}}]{\RaggedRight Continuous, per-iteration adaptation; optional rolling repack.} \\

\makecell[{{p{3.1cm}}}]{Decision basis} &
\makecell[{{p{3.6cm}}}]{\RaggedRight Performance estimation or ILP objectives.} &
\makecell[{{p{3.6cm}}}]{\RaggedRight Fairness / throughput objectives.} &
\makecell[{{p{3.8cm}}}]{\RaggedRight Combined job \& system scoring (utility + global policy).} \\
\bottomrule
\end{tabular}
\end{table}

\section{Algorithmic Design and Scheduling Policy}
\label{sec:algorithm}
This section provides a more detailed analysis of several key aspects of the \gls{JASDA} concept introduced in Section~\ref{sec:conceptual_framework}. It defines the scoring model, the clearing procedure, and presents pseudocode for a single scheduling iteration, representing one complete \gls{JASDA} interaction cycle over an announced window. Complexity considerations are also discussed.

\subsection{Variant Generation and Eligibility}
\subsubsection*{Notation and Assumptions}
Consider a set of jobs $\mathcal{J} = \{J_1, J_2, \ldots, J_n\}$ and a set of \glspl{MIG} slices $\mathcal{S} = \{s_1, s_2, \ldots, s_K\}$, each with a fixed memory capacity $c_k$ and a time-dependent availability pattern. Scheduling occurs over a finite time horizon divided into a sequence of \emph{iterations}, during which the scheduler advertises one available time–capacity window~$w^\ast$ (as defined in Section~\ref{subsec:announcement_phase}). 

Each job $J_i$ maintains its internal computational structure and \gls{FMP} describing its expected memory consumption over time. Jobs may be paused, resumed, or decomposed into multiple subjobs; however, individual subjobs are treated as non-preemptive execution blocks that must complete once scheduled, ensuring atomicity within each announced window.

To avoid excessive temporal fragmentation and unnecessary overhead, the scheduler enforces a global lower bound on subjob duration, denoted as $\tau_{\min} > 0$. This constraint ensures that each scheduled subjob occupies its assigned slice for a sufficiently long interval to justify the scheduling and activation costs, thereby preventing energy-inefficient short bursts of computation (thrashing).

\medskip
Unless otherwise noted, the following assumptions hold throughout this section:

\begin{enumerate}[label=\textbf{(A\arabic*)}, leftmargin=2em]
    \item \textbf{Static slice capacity.}  
    Slice capacities $c_k$ remain constant during the scheduling horizon; no dynamic slice reconfiguration is performed within a single iteration.

    \item \textbf{Job independence.}
    All jobs are considered independent agents, meaning that their execution and bidding decisions are made autonomously without inter-job dependencies or coordination. This assumption simplifies the per-window clearing process, as the scheduler only needs to ensure temporal and capacity non-conflicts across slices. Extensions to data-reuse or affinity-aware variants are left for future work.

    \item \textbf{Discrete scheduling iterations.}  
    The scheduler operates in discrete iterations; in each iteration, one window~$w^\ast$ is announced and corresponding bids (variants) are collected.
\end{enumerate}

\subsubsection*{Subjob Eligibility and Safe-by-Construction Property}
As introduced in Section~\ref{subsec:variant_generation}, each job may generate one or more \emph{variants} $v_{i,k,w^\ast}$ representing candidate subjobs within the announced window $w^\ast$. A variant is considered \emph{eligible} if it satisfies both of the following conditions:

\begin{enumerate}[label=\textbf{(\alph*)}, leftmargin=2em]
    \item \textbf{Probabilistic safety (safe-by-construction).}  
    Given the job’s \gls{FMP}, the probability of exceeding capacity~$c_k$ at any time over the predicted execution interval must be bounded by~$\theta$:
    \[
    \Pr\!\left(\max_{t \in [t_{\mathrm{start}},\, t_{\mathrm{start}}{+}\tilde{\Delta t}_i]} \mathrm{RAM}_i(t) > c_k \;\middle|\; \mathrm{FMP}_i\right) \le \theta.  
    \]
    This guarantees that only subjobs \emph{safe-by-construction} are exposed to the scheduler.

    \item \textbf{Slice-specific constraints.}  
    Additional requirements may apply, e.g., affinity to the same slice as a predecessor for data reuse, or locality constraints tied to memory controllers or interconnect topology.
\end{enumerate}

Only variants satisfying both conditions are submitted as bids. Multiple eligible variants may be generated per iteration, possibly overlapping in time; the scheduler later enforces global feasibility when selecting a temporally non-conflicting subset.

\subsubsection*{Outcome of the Generation Phase}
At the end of each scheduling iteration, jobs may provide a set of \emph{eligible}, locally scored variants (using local utility function $h(v_{i,k,w^\ast})$) that are submitted to the scheduler for evaluation. Participation is optional---jobs respond only if at least one \emph{eligible variant} exists for the currently advertised window~$w^\ast$. If no such variant can be generated, the job remains silent during that iteration. The union of all submitted variants thus defines the feasible bidding space of the iteration, from which the scheduler subsequently computes the \emph{global utility score} and performs the final selection in accordance with the global optimization policy.

\subsection{Scoring Model}
\label{subsec:scoring_model}
As introduced in Section~\ref{sec:conceptual_framework} (Step~4), each variant \(v_{i,k,w^\ast}\)---representing a specific subjob of job \(J_i\) assigned to slice \(s_k\) within the announced scheduling window \(w^\ast\)---is associated with a composite \emph{score} that balances job-side and system-side objectives. For brevity, we write \(v\) whenever the context makes the triplet \((i,k,w^\ast)\) implicit.

\medskip
The scheduler evaluates a convex combination of job- and system-side utilities:
\begin{equation}
\label{eq:score_raw}
\mathrm{Score}_{\text{raw}}(v_{i,k,w^\ast}) 
\;=\; \lambda\, h(v_{i,k,w^\ast}) 
\;+\; (1{-}\lambda)\, f_{\mathrm{sys}}(v_{i,k,w^\ast}), 
\qquad \lambda\in[0,1].
\end{equation}

\paragraph{Job-side utility.}
Each job attaches a local utility \(h(v)\in\mathbb{R}\) to its variant, reflecting job–specific preferences such as expected \gls{JCT} reduction, \gls{QoS} adherence, 
or estimated energy consumption. Here, \emph{energy} refers to the predicted cost of executing the subjob---for instance, based on historical GPU power profiles 
or runtime estimations of compute intensity and memory access patterns. This component enables jobs to express energy-aware preferences, discouraging variants that would lead to disproportionately high energy use relative to their expected performance gain. Several $h(v)$ features may be \gls{TRP}/\gls{FMP}-derived (e.g., predicted remaining runtime quantile, expected stall risk, or memory headroom). This allows jobs to express preferences that align with their probabilistic execution profiles.

\paragraph{System-side utility.}
The scheduler complements the job-side component with a system contribution \(f_{\mathrm{sys}}(v)\), representing global objectives such as utilization gain, slack reduction, or defragmentation improvement.

\vspace{1em}
\noindent
To formalize the scoring process, both job-side and system-side utilities can be expressed as weighted combinations of interpretable factors:
\begin{align}
h(v) &= \sum_{i=1}^{n} \alpha_i \, x_i(v), \\
f_{\mathrm{sys}}(v) &= \sum_{j=1}^{m} \beta_j \, y_j(v)
\end{align}
where $x_i(v)$ and $y_j(v)$ represent the $i$-th and $j$-th scoring features of the job- and system-side utilities, respectively, and $\alpha_i, \beta_j \ge 0$ are policy-dependent weights.

\paragraph{Normalization and non-negativity.}
To ensure comparability across heterogeneous metrics and maintain scale consistency, both utility functions are normalized to the unit interval so that \(\mathrm{Score}(v)\in[0,1]\) for all eligible variants.
Each scoring feature is transformed into a normalized feature \(\phi_i(v)\) or \(\psi_j(v)\), defined on the unit interval and oriented so that higher values consistently indicate higher desirability:
\[
\tilde h(v)=\sum_{i} \alpha_i \,\phi_i(v),\qquad
\tilde f_{\mathrm{sys}}(v)=\sum_{j} \beta_j \,\psi_j(v),
\]
where each feature \(\phi_i,\psi_j\in[0,1]\)  and the weights satisfy \(\sum_i \alpha_i\le 1\), \(\sum_j \beta_j\le 1\).
Typical transformations include:
\[
\begin{aligned}
\phi_{\mathrm{JCT}}(v) &= 1 - \frac{\Delta\mathrm{JCT}(v)}{\Delta\mathrm{JCT}_{\max}}, \\[4pt]
\phi_{\mathrm{QoS}}(v) &= \mathbf{1}[\text{meets QoS}], \\[4pt]
\psi_{\mathrm{energy}}(v) &= 1 - \frac{E(v)}{E_{\max}}, \\[4pt]
\psi_{\mathrm{mem\_headroom}}(v) &= \mathbb{E}\!\left[\frac{c_k - \mathrm{RAM}_i(t)}{c_k}\right]_{t \in I(v)}.
\end{aligned}
\]
ensuring nonnegativity by construction.

\medskip
The normalized composite score, used during the clearing and selection phase, is then given by:
\begin{equation}
\label{eq:score_normalized}
\mathrm{Score}(v_{i,k,w^\ast}) 
\;=\; 
\lambda\,\tilde h(v_{i,k,w^\ast})
+ (1{-}\lambda)\,\tilde f_{\mathrm{sys}}(v_{i,k,w^\ast}),
\qquad 
\lambda\in[0,1].
\end{equation}
This normalized form serves as the operative measure for evaluating and comparing variants across jobs, slices, and scheduling windows in subsequent optimization stages.

The top-level trade-off between job-centric and system-centric objectives is governed by the policy weight~$\lambda$, while the coefficients~$\alpha_i$ and~$\beta_j$ control the relative emphasis of individual features within each component. Table~\ref{tab:lambda_effects} summarizes representative settings of~$\lambda$ and their qualitative effects.

\begin{table}[H]
\centering
\caption{Illustrative effects of the policy parameter~$\lambda$}
\label{tab:lambda_effects}
\begin{tabularx}{\linewidth}{l c >{\raggedright\arraybackslash}X}
\toprule
\textbf{Policy Type} & $\boldsymbol{\lambda}$ & \textbf{Qualitative Effect on Scheduling Behavior} \\
\midrule
QoS-first         & 0.7 & Scheduler prioritizes job-centric metrics such as latency, fairness, and QoS adherence, possibly at the expense of overall utilization. \\
Balanced          & 0.5 & Balanced consideration between job-level responsiveness and system-level efficiency. \\
Utilization-first & 0.3 & Emphasizes maximizing resource utilization and minimizing fragmentation, even if individual job latency increases. \\
\bottomrule
\end{tabularx}
\end{table}

Jobs that repeatedly misreport---by systematically declaring unrealistically high or low utilities---can be gradually penalized by reducing their individual trust weight $\lambda_i$ in the composite score. This adaptive adjustment decreases the influence of unreliable jobs in the clearing process without altering their internal scoring . Together, these mechanisms help maintain a fair and stable bidding environment, ensuring consistent and self-regulating behavior of participating jobs without relying on external incentive systems.

\subsubsection{Incentives, Calibration, and Ex-Post Verification}
In the absence of explicit economic incentives, jobs in \gls{JASDA} are free to declare any self-assessed score~$\tilde{h}(v)$ for their variants. While this allows a lightweight, decentralized interaction, it also opens the possibility of strategic or inaccurate score reporting. For instance, a job might overstate its expected utility to gain scheduling priority. To preserve comparability and fairness among bids, \gls{JASDA} introduces a two-level reliability mechanism combining \emph{ex-ante calibration} and \emph{ex-post verification}. 
\textit{The following formulation represents one feasible design of such a mechanism rather than a prescriptive choice; it serves to illustrate how incentive alignment and long-term trust could be achieved in \gls{JASDA}.}

\paragraph{Ex-ante Calibration.}
During each iteration, declared scores are first calibrated through a smoothing step:
\begin{equation}
\label{eq:score_calibration}
\hat{h}(v) \leftarrow 
\gamma\, \tilde{h}(v) + (1-\gamma)\,\mathrm{HistAvg}(J_i),
\end{equation}
where $\gamma \in [0,1]$ balances the job’s current declaration and its historical reliability profile. The term $\mathrm{HistAvg}(J_i)$ represents the moving average of previously verified scores, i.e., scores validated against actual outcomes. This ensures that both sporadic and systematic over-optimistic declarations have limited long-term influence on the resulting allocation. The exact form of the moving average (e.g., simple or weighted) is left open, as its choice governs the trade-off between adaptability and stability.

\paragraph{Ex-post Verification.}
After execution, the scheduler evaluates the realized outcome of each accepted variant and compares the declared feature values within~$\tilde{h}(v)$ 
(e.g., expected runtime, memory headroom, QoS satisfaction) with their observed counterparts. \emph{Importantly, this verification step presupposes that the scheduler has access to the corresponding ground-truth measurements, i.e., that all features contributing to the job-side score can be measured or inferred from runtime observations.}
The \emph{per-feature error} is quantified as:
\begin{equation}
\epsilon_{i}(v) = 
\big| \phi_{i}(v)
- \phi_{i}^{\text{observed}}(v) \big|.
\end{equation}

To obtain a single scalar measure of \emph{per-variant error}, per-feature deviations are aggregated into a convex combination:
\[
\epsilon(v) = \sum_{n=1}^{n} w_i\, \epsilon_{i}(v),
\qquad
w_i \ge 0,\;
\sum_{i} w_i = 1.
\]
Here, $w = (w_1,\dots,w_n)$ represents a vector of non-negative feature weights that reflect their relative importance in the overall evaluation of per-variant error.
The convex form ensures interpretability and stability, as the resulting error remains bounded in $[0,1]$.

For each job~$J$, the overall reporting accuracy is evaluated over all its previously verified variants. Here, a \emph{verified variant} refers to a variant that has been executed and for which the corresponding observed feature values $\phi_i^{\mathrm{observed}}(v)$ are available, allowing ex-post comparison with the declared values.
Let $V_J^{\mathrm{verified}}$ denote the set of such variants.
The \emph{expected per-variant error} is then defined as
\begin{equation}
\mathbb{E}_v[\epsilon(v)]
\;=\;
\frac{1}{|V_J^{\mathrm{verified}}|}
\sum_{v \in V_J^{\mathrm{verified}}}
\epsilon(v),
\qquad
\mathbb{E}_v[\epsilon(v)] \in [0,1].
\end{equation}

Based on the expected per-variant error, a composite reliability metric is derived as:
\begin{equation}
\rho_J \;=\; 
\exp\!\left(-\kappa \cdot 
\mathbb{E}_v[\epsilon(v)]\right),
\qquad
\rho_J \in (0,1],
\end{equation}
where $\kappa > 0$ is a sensitivity parameter controlling how strongly deviations from declared values penalize reliability.
The exponential form ensures that reliability decays smoothly and asymptotically with increasing average error, while remaining bounded within $(0,1]$. Intuitively, $\rho_J$ acts as a \emph{trust coefficient} that down-weights the influence of unreliable jobs in subsequent calibration steps---either by reducing their score contribution or by shifting the calibration weight toward their historical baseline.

\paragraph{Feedback and Long-Term Stability.}
Each job’s reliability coefficient~$\rho_J$ is reintroduced into subsequent calibration steps to adjust the scheduler’s trust in its future declarations. 
For instance, the calibrated job-side utility can be updated as:
\[
\hat{h}(v) \leftarrow 
\rho_J\, \tilde{h}(v) + (1-\rho_J)\,\mathrm{HistAvg}(J),
\]
or, alternatively, $\rho_J$ can serve as a multiplicative factor applied to the entire calibrated score. Jobs that report their expected utilities consistently and accurately are gradually rewarded with higher credibility, while unreliable or opportunistic jobs experience a natural decline of influence in future iterations.

\subsection{Temporal Fairness and Age-Aware Prioritization}
\label{subsec:age_fairness}
To maintain temporal fairness and prevent starvation, \gls{JASDA} incorporates an \emph{age-aware scoring adjustment} that gradually increases the composite score, used by the scheduler during variant selection, of jobs whose variants have not been selected in the current scheduling iteration. 

Let $A_i(t)\in[0,1]$ denote a normalized age factor for job~$J_i$ at scheduler time~$t$, defined as a non-decreasing function of the waiting time since the the last successful scheduling of any variant of that job. The variable~$t$ represents the current iteration time within the rolling scheduling horizon. As the job’s variants remain unallocated, $A_i(t)$ increases towards~1, thereby amplifying their effective contribution to the system-side utility.

The age term is directly integrated into the existing normalized system-side scoring function~$\tilde f_{\mathrm{sys}}(v)$:
\[
\tilde f_{\mathrm{sys}}(v)
\;=\;
\sum_{j=1}^{m-1} \beta_j \,\psi_j(v)
\;+\;
\beta_{\mathrm{age}}\,A_i(t)
\]
As $A_i(t)$ grows, deferred jobs are gradually promoted in subsequent iterations. However, increasing $\tilde f_{\mathrm{sys}}(v)$ through the age factor~$A_i(t)$ does not by itself guarantee a hard upper bound on the total completion time of the job. Instead, it increases the likelihood that one or more of its variants will be admitted in future scheduling iterations, thereby reducing the expected delay and mitigating starvation in practice.

\subsection{Selection and Clearing Phase}
\label{subsec:selection_clearing}
The clearing phase determines which of the submitted variants are selected for execution within the announced window $w^\ast$. All candidate variants $v_{i,k,w^\ast}$ associated with slice $s_k$ compete for allocation within the same temporal execution interval, subject to the following constraints:
\begin{enumerate}[label=(\roman*), leftmargin=2em]
    \item no two selected variants may overlap in time on the same slice $s_k$; and
    \item each selected variant must be \emph{eligible} according to the definition provided in Section~\ref{subsec:variant_generation}
\end{enumerate}

The scheduler aggregates all variants submitted by jobs and evaluates them using the normalized composite scoring function defined in Eq.~(\ref{eq:score_normalized}). The system-side score may include an age term (Section~\ref{subsec:age_fairness}) to prioritize long-waiting jobs.
Let $V_i$ denote the set of all candidate variants $v_{i,k,w^\ast}$ associated with job~$J_i$.
The resulting pool 
\[
V = \bigcup_{J_i \in \mathcal{J}} V_i
\]
constitutes an instance of the \emph{\gls{WIS}} problem, where each variant $v_{i,k,w^\ast}$ represents an interval $[t_{\mathrm{start}}, t_{\mathrm{end}}]$ with an associated weight $\mathrm{Score}(v_{i,k,w^\ast})$.

An optimal solution $\widehat{S} \subseteq V$ is obtained by selecting a set of pairwise temporally non-overlapping variants that maximizes the total composite score within the announced window $w^\ast$. The complete per-window decision procedure is summarized in Algorithm~\ref{alg:JASDA_iteration}, which illustrates a single scheduling iteration over window $w^\ast$. During each iteration, the scheduler collects all submitted variants, computes their composite scores, and applies the optimal \gls{WIS}-based selection algorithm to determine $\widehat{S}$.

\begin{algorithm}[t]
\caption{JASDA: Single-window scheduling iteration on $w^\ast=(s_k, c_k, t_{\min}, \Delta t)$}
\label{alg:JASDA_iteration}
\begin{algorithmic}[1]
\State \textbf{Input:} window $w^\ast$, job set $\mathcal{J}$, policy parameters $(\lambda, \alpha_1\!:\!\alpha_n, \beta_1\!:\!\beta_m, \tau_{\min})$
\State \textbf{Output:} committed subjob set $\widehat{S}$ within $w^\ast$
\vspace{3pt}
\State \textbf{Announce} $w^\ast$ to all $J_i \in \mathcal{J}$
\ForAll{$J_i \in \mathcal{J}$}
    \State $V_i \leftarrow$ \textsc{GenerateVariants}$(J_i, w^\ast, \tau_{\min})$ 
    \ForAll{$v \in V_i$}
        \State compute $ \tilde h(v)$ and $ \tilde f_{\mathrm{sys}}(v)$ \Comment{features may query TRP/FMP over $I(v)$}
        \State $\mathrm{Score}(v) \leftarrow \lambda\, \tilde h(v) + (1{-}\lambda)\, \tilde f_{\mathrm{sys}}(v)$ \Comment{Eq.~(\ref{eq:score_normalized})}
    \EndFor
\EndFor
\vspace{3pt}
\State $V \leftarrow \bigcup_{J_i} V_i$
\State $\widehat{S} \leftarrow$ \textsc{SelectBestCompatibleVariants}$(V, \mathrm{Score})$ \Comment{optimal weighted interval scheduling}
\vspace{3pt}
\State \textsc{Commit} $\widehat{S}$ on slice $s_k$; update layout and job statistics
\end{algorithmic}
\end{algorithm}

\paragraph{Selection routine.}
Since all variants correspond to the same time–capacity window $[t_{\min},\, t_{\min}{+}\Delta t]$ on slice $s_k$, the selection phase reduces to an instance of the \emph{\gls{WIS}} problem. The scheduler seeks to identify a maximum-score subset $\widehat{S} \subseteq V$ of pairwise temporally non-overlapping variants, maximizing the total normalized composite score defined in Eq.~(\ref{eq:score_normalized}).

The \textsc{SelectBestCompatibleVariants} routine implements the classical dynamic programming formulation of the \gls{WIS} problem, executed after sorting all candidate variants by their end times. This guarantees an optimal subset under a sum-based scoring model and achieves $\mathcal{O}(M \log M)$ complexity for $M = |V|$. 

Once the clearing process for $w^\ast$ is finalized, the selected subjobs are committed for execution, and both the slice layout and internal job statistics are updated accordingly. Through this iterative cycle of window announcement, variant generation, scoring, and clearing, the \gls{JASDA} continuously adapts to evolving workload conditions, incrementally filling idle gaps and improving the overall utilization of the \gls{MIG} device over time.

\subsection{Implementation Example}
\label{subsec:implementation_example}
To illustrate the algorithmic flow of the proposed scheduling algorithm, his section presents a simplified and deterministic example of one scheduling iteration performed on a single slice. For clarity, start and end times are treated as fixed values, although in practice they are predicted probabilistically from the job’s \gls{TRP}/\gls{FMP} models. The goal of this example is to show how the scheduler announces a time--capacity window, how jobs respond with eligible and scored subjob variants, and how the final selection (clearing) is made.

\paragraph{Step 1: Window announcement.}
At time $t_{\min}=40$, the scheduler detects an available window on slice $s_2$ with capacity $c_2 = 20\,\mathrm{GB}$ and duration $\Delta t = 10$.
It therefore announces the window
\[
w^\ast = (s_2, c_2, t_{\min}=40, \Delta t = 10)
\]
to all jobs currently in the set $\mathcal{J} = \{J_A, J_B, J_C\}$.

\paragraph{Step 2: Variant generation and scoring.}
Each job evaluates the announced window and decides whether to generate subjob variants that can safely execute within it. Job $J_A$ proposes two variants differing in duration; $J_B$ offers one longer variant; $J_C$ generates no variant, as its next feasible preemption point lies beyond $t_{\min}{+}\Delta t$.

The submitted variants and their associated scores are summarized in Table~\ref{tab:variant_example}. Job-side utilities $ \tilde h(v_i)$, system-side contributions $ \tilde f_{\mathrm{sys}}(v_i)$, and the final scores $\mathrm{Score}(v_i)$ are computed according to Eq.~(\ref{eq:score_normalized}) with $\lambda=0.6$.

\begin{table}[h]
\centering
\caption{Example of subjob variants submitted for window $(s_2, c_2 = 20\,\mathrm{GB}, t_{\min}=40, \Delta t = 10)$ and their scores.}
\label{tab:variant_example}
\begin{tabular}{@{}lcccccc@{}}
\toprule
\textbf{Job} & \textbf{Variant ID} & \textbf{Start} & \textbf{End} & $ \tilde h(v_i)$ & $ \tilde f_{\mathrm{sys}}(v_i)$ & $\mathrm{Score}(v_i)$ \\ 
\midrule
$J_A$ & $v_{A1}$ & 40 & 47 & 0.75 & 0.55 & 0.67 \\
$J_A$ & $v_{A2}$ & 47 & 50 & 0.60 & 0.70 & 0.64 \\
$J_B$ & $v_{B1}$ & 40 & 50 & 0.80 & 0.60 & 0.72 \\
$J_C$ & --- & --- & --- & --- & --- & --- \\
\bottomrule
\end{tabular}
\end{table}

\paragraph{Step 3: Selection and commitment.}
The scheduler aggregates all submitted variants into a single candidate pool $V = \{v_{A1}, v_{A2}, v_{B1}\}$ and invokes the selection routine (\textsc{\SelectBestCompatibleVariants{}}) described in Algorithm~\ref{alg:JASDA_iteration}.
Because $v_{B1}$ overlaps with both $v_{A1}$ and $v_{A2}$, the \gls{DP} selector compares the total scores of compatible subsets and chooses the optimal one:
\[
\widehat{S} = \{v_{A1}, v_{A2}\}, \quad \text{with total score } \sum_{v \in \widehat{S}} \mathrm{Score}(v) = 1.31.
\]
The chosen variants $v_{A1}$ and $v_{A2}$ are committed to slice $s_2$, while $v_{B1}$ is deferred to a later scheduling iteration.

\paragraph{Step 4: Outcome and interpretation.}
This example demonstrates how \gls{JASDA} balances local job preferences and system-level objectives within a single iteration. Job $J_A$ successfully occupies the full available window by submitting multiple short variants, while $J_B$’s longer subjob is not selected in this iteration and may be resubmitted in response to future window announcements. Such interactions have the potential to yield to high utilization with minimal fragmentation, while preserving fairness in access to limited slice resources. Because the clearing process operates in a rolling manner, jobs may generate and resubmit new variants in response to emerging windows, enabling continuous adaptation.

\subsection{Computational Complexity and Theoretical Properties}
\label{subsec:complexity_properties}
The computational complexity of the \gls{JASDA} scheduling algorithm is dominated by the variant generation and selection phases, which together define the total cost of each scheduling iteration.

\paragraph{Per-iteration complexity.}
Let $\mathcal{J}$ denote the set of jobs participating in the current iteration. 
Each job $J_i \in \mathcal{J}$ may submit a finite set of eligible variants 
\[
V_i = \{\, v_{i,k,w^\ast}^{(1)},\, v_{i,k,w^\ast}^{(2)},\, \ldots \,\},
\]
where each variant $v_{i,k,w^\ast}^{(j)}$ represents a distinct feasible execution option within the currently announced window $w^\ast$.
The union of all submitted variants forms the candidate pool
\[
V = \bigcup_{J_i \in \mathcal{J}} V_i,
\]
and we denote its total size by $M = |V|$.
\begin{itemize}
    \item \textbf{Variant generation.}  
    Each job $J_i$ locally constructs a set of \emph{eligible variants} based on its current internal state and the parameters of the announced window $w^\ast$.  
    This procedure may depend on multiple job-specific factors---such as progress within the parent job, synchronization boundaries, or data exchange points---and therefore cannot be assumed constant over time. Let $t_{\mathrm{gen}}$ denote the average computational time required to generate one eligible variant.  
    The aggregate job-side cost of this step is thus proportional to the total number of generated variants, yielding a complexity of $\mathcal{O}(M) \cdot t_{\mathrm{gen}}$ for $M = \sum_i |V_i|$ variants across all jobs.

    \item \textbf{Selection.}  
    The \textsc{SelectBestVariants} routine solves a \gls{WIS} instance defined on the candidate pool $V$. Sorting all variants by end time and computing the optimal temporally non-overlapping subset via \gls{DP} yields a time complexity of $\mathcal{O}(M \log M)$ per window. This guarantees an optimal selection under sum-based scoring model while maintaining scalability for large variant sets.
\end{itemize}

\medskip
Consequently, the overall per-iteration complexity of \gls{JASDA} is given by
\[
\mathcal{O}(M) \cdot t_{\mathrm{gen}} \;+\; \mathcal{O}(M \log M),
\]
where the first term reflects decentralized job-side computation during variant generation, and the second term corresponds to the centralized \gls{DP}-based clearing. The resulting complexity makes \gls{JASDA} applicable even in fine-grained, high-frequency scheduling regimes, provided that the average per-variant generation time $t_{\mathrm{gen}}$ remains bounded independently of system scale and workload size.

\paragraph{Asymptotic properties.}
Assuming that job arrivals follow a stochastic process with bounded rate $\lambda_{\mathrm{arr}}$ and that each job can submit at most $V_{\max}$ variants per iteration, the expected scheduling overhead per unit time is asymptotically given by
\[
\mathcal{O}\!\left(
\lambda_{\mathrm{arr}}\,V_{\max}\,
\big(t_{\mathrm{gen}} + \log(\lambda_{\mathrm{arr}} V_{\max})\big)
\right).
\]

The first component represents the decentralized job-side cost of variant generation, while the logarithmic term corresponds to the centralized clearing step performed by the scheduler. As long as the average per-variant generation time $t_{\mathrm{gen}}$ remains bounded and $V_{\max}$ does not scale with the system size, 
the total scheduling overhead grows \emph{quasi-linearly} with the job arrival rate and remains unaffected by workload heterogeneity. This follows from the fact that the overall complexity depends only on the total number of submitted variants, not on the individual characteristics or variability of the jobs. Consequently, \gls{JASDA} maintains expected scalable and stable runtime behavior even in dynamic, multi-slice environments with diverse workloads.

\paragraph{Optimality and asymptotic suboptimality.}
For each advertised window $w^\ast$, the weighted interval selection step in \gls{JASDA} yields an optimal subset of temporally non-overlapping variants with respect to the composite scoring function $\mathrm{Score}(v_{i,k,w^\ast})$. Hence, \gls{JASDA} achieves \emph{local optimality per scheduling iteration}. However, since each iteration is evaluated myopically---based only on the currently visible scheduling window---the cumulative schedule produced over time cannot be globally optimal in hindsight, i.e., relative to an \emph{offline clairvoyant scheduler} with full future knowledge.

Classical results from online optimization and auction theory show that under stationary arrival processes and \emph{nonnegative, bounded, and monotone} utility functions, decentralized iterative allocation algorithms can achieve predictable performance within a constant factor of the offline optimum~\cite{blum2006,buchbinder2007,chin2006,gunther2013}. These findings motivate, but do not strictly determine, the theoretical behavior of \gls{JASDA}. The proposed scheduling mechanism differs from standard market-clearing and primal–dual schemes in two key aspects: \\
(i) its scoring model combines job-side utility with system-side objectives such as utilization gain and temporal fairness, and \\ 
(ii) its iterative clearing step integrates the \emph{temporal fairness mechanism} (Section~\ref{subsec:age_fairness}), which mitigates starvation and promotes bounded \emph{expected} waiting times by gradually increasing the score of deferred job \\ 

While a formal competitive bound for \gls{JASDA} remains an open question, its design follows the same structural principles that yield bounded suboptimality in classical online allocation and market-clearing algorithms. Under stationary arrivals and monotone scoring, the iterative clearing dynamics are expected to maintain asymptotic stability and empirically limited deviation from the offline optimum, as observed in analogous myopic bidding systems.

\section{Discussion}
\label{sec:illustrative_example}

The design philosophy of \gls{JASDA} can be summarized through several inherent operational principles:

\begin{enumerate}[label=(\alph*)]
    \item \textbf{Autonomy and locality.}  
    Each job autonomously decides which slices and time segments to target, based on its own objectives and constraints. No centralized coordination is required; jobs interact with the scheduler only by responding to its window announcements.
    
    \item \textbf{Safe-by-construction subjobs.}  
    All submitted variants are \emph{a priori} eligible, providing probabilistic guarantees that no slice capacity violation can occur once accepted.

    \item \textbf{Local auction dynamics.}  
    The scoring mechanism effectively forms a one-shot auction: within each scheduling iteration, subjobs ``bid'' for execution slots by submitting their scored variants, and the scheduler acts as an auctioneer performing local clearing to select an optimal, temporally non-overlapping subset.
    
    \item \textbf{Elasticity and adaptivity.}  
    Since unselected variants simply expire, each job can locally adjust its future variant proposals in response to newly announced windows---for instance, by generating shorter subjobs when offered windows are limited in duration, or by targeting different slice sizes as availability changes. This adaptivity arises purely from local information provided by the scheduler, without requiring any global system knowledge or lookahead capability.

    \item \textbf{Temporal fairness and long-term stability.}  
    The scheduling mechanism incorporates age-aware adjustments that gradually increase the effective score of deferred jobs, thereby improving their chances of selection in subsequent iterations. The gradual age-aware adjustment maintains overall scheduling efficiency by preventing abrupt change in relative scores, while its damping effect promotes long-term stability by reducing starvation and oscillations under dynamic and heterogeneous workloads.

    \item \textbf{Transparency and verifiability.}  
    All declared scores and subsequent scheduling outcomes are subject to post-hoc verification against measurable execution data. This transparency fosters accountability and prevents strategic misreporting, as each job’s historical reliability directly influences future calibration. By design, this feedback loop promotes trust and self-regulation within the scheduling system without the need for explicit external enforcement.

    \item \textbf{Scalability.}  
    The mechanism exhibits quasi-linear scalability with respect to the number of slices per GPU, as each slice window can be planned independently within a unified scheduling cycle. When extended to multi-GPU or heterogeneous environments, scalability remains conceptually preserved, but practical efficiency depends on the cost of inter-GPU data movement and synchronization. In such settings, subjob migration across GPUs may incur non-negligible latency or energy overhead, suggesting that global coordination policies should favor locality whenever possible.
    
\end{enumerate}

\noindent
\textbf{\gls{JASDA} transforms static scheduling into an adaptive, self-regulating ecosystem. Through decentralized decisions, transparent scoring, and verifiable feedback, the system achieves scalable coordination, bounded variability, and long-term fairness --- not by pursuing global optimum, but by allowing optimal behavior to emerge from local interactions.}

\subsection{Challenges and Open Issues}
\label{subsec:challenges}
While the \gls{JASDA} provides a conceptually clean and theoretically grounded approach to job–aware scheduling, several practical and theoretical challenges remain. These open issues highlight the need for further refinement before the concept can evolve into a fully deployable scheduling paradigm for production-level \gls{MIG}-enabled environments.

\paragraph{(a) Timeliness of subjob generation.}
A fundamental challenge concerns the latency of subjob generation. If jobs fail to produce eligible variants quickly enough, the scheduler may issue announcements without receiving a sufficiently diverse or competitive set of bids.  This sparsity degrades overall scheduling quality, as the candidate pool $V$ becomes too limited for effective selection.  
Two possible mitigations are being investigated: \\ 
(i) introducing a time offset between the announcement of an upcoming window and its effective start, granting jobs additional time for variant preparation; and \\ 
(ii) allowing jobs to pre-generate tentative variants for future windows, trading adaptivity for responsiveness.  \\
Both directions require empirical evaluation to quantify their impact on overall performance and system responsiveness.

\paragraph{(b) Fairness and policy calibration.}
Although the scoring mechanism provides explicit parameters $(\lambda, \alpha_1{:}\alpha_n, \beta_1{:}\beta_m)$ to control fairness–efficiency trade-offs, practical calibration of these parameters in multi-tenant or multi-user environments is non-trivial. Developing principled or learning-based methods for policy tuning represents an important direction for future work, especially to ensure long-term fairness and predictability across heterogeneous workloads.

\paragraph{(c) Adaptive window selection.}
The effectiveness of the announcement phase depends critically on which idle scheduling windows are advertised for bidding. The current \gls{JASDA} prototype prioritizes announcing windows with the earliest start times, minimizing latency between announcement and subjob generation. Alternative strategies, such as slack-aware, fragmentation-aware, or learning-based policies, may yield better performance under dynamic workloads. A deeper investigation into these strategies---and their interaction with the probabilistic safety layer---remains an open and promising line of research.

\paragraph{(d) Information visibility and transparency.}
The current \gls{JASDA} design assumes that jobs have access only to the scheduling windows explicitly announced by the scheduler. While this ensures decentralization and simplicity of design, it also limits each job’s situational awareness and ability to adjust its strategy based on the global scheduling state. Future extensions may explore controlled transparency models---such as partial schedule visibility or aggregated state indicators---that provide richer feedback without compromising scalability or fairness. Determining the optimal degree of information sharing between the scheduler and participating jobs thus remains an open and highly relevant research question.

\paragraph{(e) Cooperation among jobs.}
The current \gls{JASDA} formulation assumes purely non-cooperative behavior, where each job independently reacts to the scheduler’s announcements without direct coordination with others. This design preserves decentralization but it may also limit the system’s ability to exploit synergistic opportunities --- for example, mutually adjusting subjob variants to reduce fragmentation or balance load across slices. 

Allowing controlled forms of cooperation or information sharing \emph{among jobs} could improve global utilization and stability, but also raises new questions of incentive compatibility, fairness, and complexity. This issue is closely related to the previous issue of information visibility and transparency and defines an open research direction at the intersection of distributed optimization and game-theoretic scheduling.

\paragraph{(f) Implementation pathway.}
Finally, realizing the \gls{JASDA} concept in practice requires a robust runtime layer supporting bid–response communication between jobs and the scheduler.  
A prototype implementation of this bidirectional protocol, together with automated subjob-generation routines, is currently being prepared within the \emph{R} ecosystem. This prototype will form the foundation for an upcoming companion paper focusing on empirical evaluation, runtime integration, and system-level deployment.

\medskip
Addressing these challenges will be key to establishing \gls{JASDA} as a viable foundation for intelligent, data-driven scheduling in \gls{MIG}-enabled heterogeneous computing environments.

\section{Conclusion and Outlook}
\label{sec:conclusion}
This paper introduced \gls{JASDA}, a novel concept for dynamic scheduling on \gls{MIG}-enabled \gls{GPU}s. Departing from centralized optimization and static job descriptors, \gls{JASDA} decomposes scheduling into iterative \emph{bid–clear} cycles where jobs actively participate through declarative subjob proposals and policy-guided selection. This design redefines scheduling as a distributed interaction process rather than a single optimization task.

\medskip
\noindent
\textbf{Key contributions.}
\gls{JASDA} demonstrates that:
\begin{itemize}
    \item \textbf{Dynamic job participation.}  
    Scheduling can be formalized as a cyclic, decentralized negotiation process with bounded computational overhead, allowing jobs to actively participate in resource allocation without central coordination.

    \item \textbf{Probabilistic safety-by-construction.}  
    Eligibility constraints ensure that all submitted subjob variants respect slice capacities with high confidence, preventing probabilistic capacity violations or runtime interference.

    \item \textbf{Policy-driven scoring.}  
    The scoring model exposes explicit control parameters that allow fine-grained balancing between fairness, utilization, and temporal responsiveness.

    \item \textbf{Internal score verification.}  
    The concept supports runtime monitoring and cross-validation of declared versus observed subjob scores, enabling adaptive trust calibration and robustness against misreporting.

    \item \textbf{Temporal prioritization.}  
    Deferred jobs are gradually promoted through an age-aware adjustment mechanism that increases their composite score over time, enhancing temporal fairness and mitigating starvation.
\end{itemize}

\medskip
Beyond its algorithmic clarity, \gls{JASDA} offers a unifying perspective that connects online optimization, auction theory, and resource-aware scheduling. 
It thus represents a shift from static, scheduler-centric control toward a self-regulating scheduling ecosystem driven by intentional job behavior and transparent system feedback.

\paragraph{Future directions.}
Building upon this foundation, several research directions emerge:
\begin{enumerate}[label=(\alph*)]
    \item \textbf{Empirical validation.}  
    A forthcoming study will benchmark \gls{JASDA} against established schedulers such as \textsc{Pollux}, \textsc{HiveD}, and \textsc{MISO}, evaluating utilization, fairness, and responsiveness under diverse workload mixes.

    \item \textbf{Learning-augmented policies.}  
    Integrating machine learning into the scoring functions $\tilde h(v)$ and $ \tilde f_{\mathrm{sys}}(v)$ may enable context-aware adaptation and predictive variant evaluation, further improving both responsiveness and robustness in dynamic environments.

    \item \textbf{Policy calibration and fairness.}  
    While \gls{JASDA} exposes explicit parameters $(\lambda, \alpha_i, \beta_j)$ to balance fairness and efficiency, practical tuning in multi-tenant environments remains non-trivial. Future work will explore principled and learning-based calibration methods to ensure long-term temporal fairness and predictability across heterogeneous workloads.

    \item \textbf{Information visibility and cooperative adaptation.}  
    The current design assumes that jobs only observe the scheduler’s announced windows. Extending this model toward controlled transparency or partial cooperation among jobs may unlock higher utilization and stability, but also raises open questions about incentive compatibility and distributed decision-making.

    \item \textbf{Runtime implementation.}  
    A prototype of the bidirectional job scheduling protocol and automated subjob generation routines is currently being prepared within the ecosystem \emph{R}, which outlines a practical path toward runtime deployment and empirical evaluation.
\end{enumerate}

\medskip
\noindent
Together, these directions outline a research agenda toward the next generation of self-adaptive, explainable, and data-driven scheduling frameworks for heterogeneous \gls{MIG}-enabled clusters. By coupling decentralized job intelligence with verifiable feedback and policy-level transparency, \gls{JASDA} sets the stage for schedulers that not only react to workloads but also \emph{learn, reason, and negotiate} --- bridging the gap between theoretical optimality and real-world deployability.

\clearpage
\printnoidxglossaries
\clearpage


\bibliographystyle{unsrturl}

\end{document}